%% file: main.tex
\documentclass[twocolumn,english]{aastex6}
\usepackage[T1]{fontenc}
\usepackage[latin9]{inputenc}
\usepackage{units}
\usepackage{bm}
\usepackage{amstext}
\usepackage{amssymb}
\usepackage{graphicx}

\makeatletter
\usepackage{amsmath}
\usepackage{gensymb}

\hypersetup{allcolors=blue}

\journalinfo{The Astronomical Journal}

\makeatother

\usepackage{babel}
\begin{document}

\title{A Long-term photometric variability and spectroscopic study of luminous blue variable {\sl AF And} in M31}

\shorttitle{Long-term photometric variability and spectroscopic study of {\sl AF And}}

\author{Yogesh C. Joshi\altaffilmark{1}}

\author{Kaushal Sharma\altaffilmark{2}}

\author{Anjasha Gangopadhyay\altaffilmark{1}}

\author{Rishikesh Gokhale\altaffilmark{3}}

\author{Kuntal Misra\altaffilmark{1}}

\email{yogesh@aries.res.in}

\affil{$^{1}$ Aryabhatta Research Institute of Observational Sciences (ARIES), Manora peak, Nainital 263002, India\\
$^{2}$ Inter-University Centre for Astronomy and Astrophysics (IUCAA), Post Bag 4, Pune, Maharashtra 411007\\
$^{3}$ Department of Physics and Astronomy, Stony Brook University, Stony Brook, NY 11794-3800, USA}
\begin{abstract}
We present photometric and spectroscopic analysis of the Hubble Sandage variable {\sl AF And} in M31. The data has been taken under the Nainital Microlensing Survey during 1998-2002 and follow-up observations were carried out until 2011. During this period, photometric observations in Cousins $R$ and $I$ bands were obtained for 169 nights spanning over about 5000 days. {\sl AF And} has shown a prominent outburst around mid-January in 1999 followed by a gradual decrease in brightness of about 1.5 mag in the next 3 years with a declining rate of $\sim$0.0015 mag~day$^{-1}$ leading to a quiescent phase at the end of 2001. After lying low for about 9 years, {\sl AF And} again went through a secondary outburst phase in late 2010 with an amplitude of 0.44 mag where it lasted for one year before fading back to its quiescence phase. The spectroscopic observations of {\sl AF And} show prominent Balmer and He I emission lines along with the comparatively weaker FeII and [FeII] emissions. Asymmetric emission line profiles in its spectrum imply the mass loss rate of about 2.2$\times$10$^{-4}$ M$_{\odot}$ yr$^{-1}$ through the stellar winds in the photosphere. Using SED fitting, we find the photospheric temperature of 33,000$\pm$3000\,K during the visual minimum. Using a weak P Cygni profile of HeI emission line, the wind terminal velocity for {\sl AF And} is found to be around 280-300 km~s$^{-1}$.
\end{abstract}

\keywords{stars: variables; stars: supergiants -- galaxies: individual: M31 - methods: data analysis}

%
\section{Introduction}\label{s:intro}
Luminous Blue Variables (LBVs) are unstable, evolved massive stars and characterized by their high intrinsic luminosities. They are generally considered to be a short but violent phase of the most massive star towards the late stage of their evolution \citep{Humphreys1994,Langer1994}. One of the distinguishing characteristics of the LBV stars is that they lie on the S Dor instability strip during the quiescent phase and contain large $L/M$ ratios \citep{Humphreys2016}. LBVs are found to exhibit multiple successive eruptions. They typically show optical variability over a wide range of amplitudes and diverse time scales \citep{Szeifert1996}. Normally outburst with amplitudes of 1 to 2 mag are observed in optical bands but sometimes, though very rare, the brightness can increase by more than 2 mag \citep{Humphreys1994,Szeifert1996,Kurtev1999}. The typical outburst phase could last from a few years to tens of years followed by an extremely long period of quiescent phase. This kind of irregular variability pattern is the hallmark of LBV phase of the most massive stars. There are two possible scenarios proposed for the existence of LBVs, one suggests that LBVs are the final stage of super-massive stars before a supernova explosion \citep{Kotak2006, Meynet2011} and second one advocates that LBVs could be a product of stellar evolution in the binary systems \citep{Smith2015}. The irregular photometric and spectroscopic variability in LBVs are related to transient eruptions, mechanism of which is still not fully understood but believed to be caused by the hydrodynamic instabilities in the stars particularly when they reach their Eddington limits of radiative stability \citep{Langer1999}. It is believed that the episodic ejection of the H envelope during the eruptions in LBV stars is a dominant mode of mass-loss for massive stars, however, strong shock waves may also play a role in some of these eruptions \citep{Smith2006,Smith2011}. In general, LBVs cover a range of objects that includes S Dor variables, P Cygni variables and Hubble-Sandage variables \citep{Conti1984}. Since only a small number ($\sim 10\%$) of super-massive stars ($M_{\sin i} > 22M_\odot$) pass through this evolutionary track \citep{Humphreys2014b}, study of LBVs helps us in understanding the evolution of such high mass stars, mass loss history and their effects on the interstellar medium.

The Hubble-Sandage variables, a subclass of LBVs, are one of the most luminous objects in the nearby spiral galaxies M31 and M33 \citep{Luyten1928,Hubble1953}. The main characteristics of these post main-sequence stars are high luminosity, blue colour, and irregular variability \citep{Hubble1953,Humphreys1975}. The sudden outburst in these stars results in an enhanced mass outflow which advances to the formation of an expanded atmosphere towards the maximum brightness phase. At this stage, the slowly expanding envelope with the speed of 100-200 km~s$^{-1}$ is relatively cool ($\sim$ 7000\,-\,9000\,K) and dense ($N \sim 10^{11}$~cm$^{-3}$) and star resembles with a very luminous F to A-type supergiant \citep{Szeifert1996}. An important photometric characteristic of LBV stars is that during an eruption phase, the bolometric luminosity remains almost constant \citep{Wolf1989,Humphreys1994}, however, the mass loss rate increases and the wind becomes opaque and cool. The change in flux variation is caused by the apparent shift in the energy distribution driven by the instability \citep{Szeifert1996}. Spectroscopically they are hot emission stars and show H, HeI, FeII and [FeII] lines with a very strong UV continuum and no Balmer jump \citep{Humphreys1978}. The high resolution spectra of these objects show the P Cygni profiles in H and HeI emission lines \citep{Gallagher1981,Humphreys2014a}. The P Cygni profile, which is characterized by strong emission lines with corresponding blue-shifted absorption lines, contains features in the stellar spectrum that points to an outflow of material in the form of either an expanding shell of gas or a strong stellar wind.

\citet{Hubble1953} discovered one such star, first of its kind in M31, {\sl AF And} (Var 19) having distinguished features of a LBV that falls in the category of Hubble-Sandage variables. A detailed analysis of this object has been carried out by many authors in the past \citep{Gallagher1981,Sharov1990,Szeifert1996,Humphreys2014b,Humphreys2017}. We also monitored this star under the Nainital Microlensing Survey that was conducted for four years during 1998-2002 in the direction of M31 \citep{Joshi2001,Joshi2005}. It should be noted here that although prime motivation of the survey was to detect gravitational microlensing events in the direction of M31, the photometric data accumulated over many years was also found to be suitable for the characterization of different class of variable stars. In earlier studies, we have presented our analysis on the microlensing event \citep{Joshi2005}, Cepheid variables \citep{Joshi2003,Joshi2010}, novae \citep{Joshi2004,Joshi2012}, and eclipsing binary stars \citep{Joshi2017}. In this study, we carry out photometric and spectroscopic analysis of the star {\sl AF And}.

Our paper is structured as follows: The past study of {\sl AF And} is briefly discussed in Section~\ref{past}. In Section~\ref{data}, we provide some information about the data used in the present analysis. The photometric study of the object is described in Section~\ref{photo}. The details about spectroscopic observations is given in Section~\ref{spec} followed by spectral analysis in Section~\ref{spec_ana}. A short discussion on our study is presented in Section~\ref{discuss} and we summarize our results in Section~\ref{conclu}.
\input{table01.tex}
\section{Previous Studies of {\sl AF And} in M31}\label{past}
There are notable photometric and spectroscopic variations of this star since its discovery by \citet{Luyten1928} and \citet{Hubble1953} and some of them are briefly discussed below.
\begin{itemize}
\item {\sl AF And} was monitored for a period of almost 36 years from 1917 to 1953 but no definite periodicity was noticed in the brightness variation of the star \citep{Hubble1953}. There were at least 5-6 major eruptions with photometric variability of more than 0.7 mag and 2-3 secondary eruptions with the visual variability of less than 0.5 mag.
\item As observed by \citet{Gallagher1981}, strong FeII emission lines and weaker [FeII] emission lines were present during the eruption phase. Further, they noted that there was absence of FeI, [FeI] and OI emission lines as well as no OII nebular lines which are normally seen in these type of stars. As indicated in \citet{Gallagher1981}, the absence of OI lines suggests the lack of extended low density envelop in the star.
\item \citet{Sharov1990} noted that there were 2 major eruptions in {\sl AF And} around 1970-1974 and 1987-1992 with photometric variability of more than 1.5 mag. They also noticed several secondary minor outbursts during those two decades.
\item \citet{Szeifert1996} studied the UV photometric and spectroscopic properties of H-S variables in M31 and M33 and derived luminosity for {\sl AF And} between $log(L/L_{\odot})$=5.9 to 6.1 and a mass loss rate of 3$\times$10$^{-5} M_\odot$ yr$^{-1}$ through IR-excess in the stellar energy distribution. Numerous FeII and [FeII] emission lines have also been observed by them along with MgII line around 2800. They also detected HeI ($\lambda$3187) along with NII (4600$<\lambda<$4643) and SiIII ($\lambda$4553,4568,4578) in UV spectrum with P Cygni profile.
\item \citet{Humphreys2014b} reported that this star has spectrum similar to an Of/late WN star with strong emission lines of NIII and HeII $\lambda$4686 and has shown very little spectroscopic variation since 2003-2004 because of its quiescence phase. \citet{Humphreys2017} also reported variation in the strength of H and HeI P-Cygni profiles but, in general, spectra of {\sl AF And} from 2010, 2013, and 2015 do not show any substantial change.
\end{itemize}
\section{Photometric Data}\label{data}
The Nainital Microlensing Survey (NMS) project was conducted for four years during 1998-2002 towards the disk of M31 where Cousins $R$ and $I$ bands photometric observations of the target field centered at $\alpha _{2000}$ = $00^{h} 43^{m} 38^{s}$; $\delta_{2000}$ = $+41^{\circ}09^{\prime}.1$ have been carried out using the 1.04-m Sampurnanand Telescope at Manora Peak, Nainital, India. The observational details of our survey can be found in \citet{Joshi2003,Joshi2010}. Although the project had been carried out with the primary aim of detecting gravitational microlensing events in the direction of M31 \citep{Joshi2001,Joshi2005}, however, photometric data was also found to be well suited for the identification of other kind of variable stars. Several surveys aimed to find microlensing events towards M31 have already unearthed thousands of variable stars as a major by-product, most of which were previously unknown \citep{Joshi2003,Ansari2004,Darnley2004,Fliri2006,Joshi2010,Lee2012,Soszy2016}. In continuation of our efforts to identify variables in the archival data taken under our survey, we searched for eruptive variables in the target field of M31. In the present study, we analyse one such variable {\sl AF And} which is present in our target field.

The Cousins $R$ and $I$ bands photometric data of {\sl AF And} was acquired on 169 nights spanning for a duration of about 5000 days which is long enough to study such kind of long-term irregular variables. In total, we collected more than 1300 frames in $R$ and $I$ bands using two different telescopes namely 1.04-m Sampurnanand Telescope (ST) and 2-m Himalayan Chandra Telescope (HCT). A log of our observations is given in Table~\ref{tbl:obs_log}. Out of 169 nights of observations, data on 149 nights were taken from 1-m ST and 20 nights were collected from 2-m HCT. It should be noted here that all the observations acquired during Nainital Microlensing Survey have been taken during October to January when M31 is visible from our site while observations before October are unlikely to happen due to monsoon season in India.

On each night, the frames in $R$ and $I$ bands have been co-added separately to increase the signal-to-noise ratio. On these co-added frames PSF-fitting photometry was performed using DAOPHOT. The typical accuracy of our photometric measurements is about 0.02 mag. A total number of 159 and 135 photometric data points were accumulated for {\sl AF And} in $R$ and $I$ bands, respectively. As our target field was not observed in both the filters on all the observing nights, the $(R-I)$ colour information could be extracted for only 126 nights. In Table~\ref{tbl:data}, we provide mid-JD of our observations (mJD), photometric magnitudes in $R$ and $I$ bands and their corresponding photometric errors. Both Tables~\ref{tbl:obs_log} and \ref{tbl:data} are available online in the complete form.
\input{table02.tex}

Basic parameters of {\sl AF And} are compiled in Table~\ref{tbl:basic_parameters}. This star has been observed photometrically in $UBVRI$ bands through a survey of local galaxies conducted by \citet{Massey2006}, in $JHK_s$ bands by 2MASS survey \citep{Skrutskie2006}, and $W_{1,2,3,4}$ bands by WISE Survey \citep{Cutri2012}. According to the latest release of GAIA mission \citep{Gaiamission}, in the DR2 release \citep{Gaiadr2}, {\sl AF And} is found to have a mean $g$-band magnitude of $17.049\pm0.011$. These photometric observations were further used to study the spectral energy distribution (SED) of the star in Section~\ref{spec_temp}.
\input{table03.tex}
\section{Photometric analysis of AF And}\label{photo}
\begin{figure*}
\includegraphics[width=17.0cm,height=11.5cm]{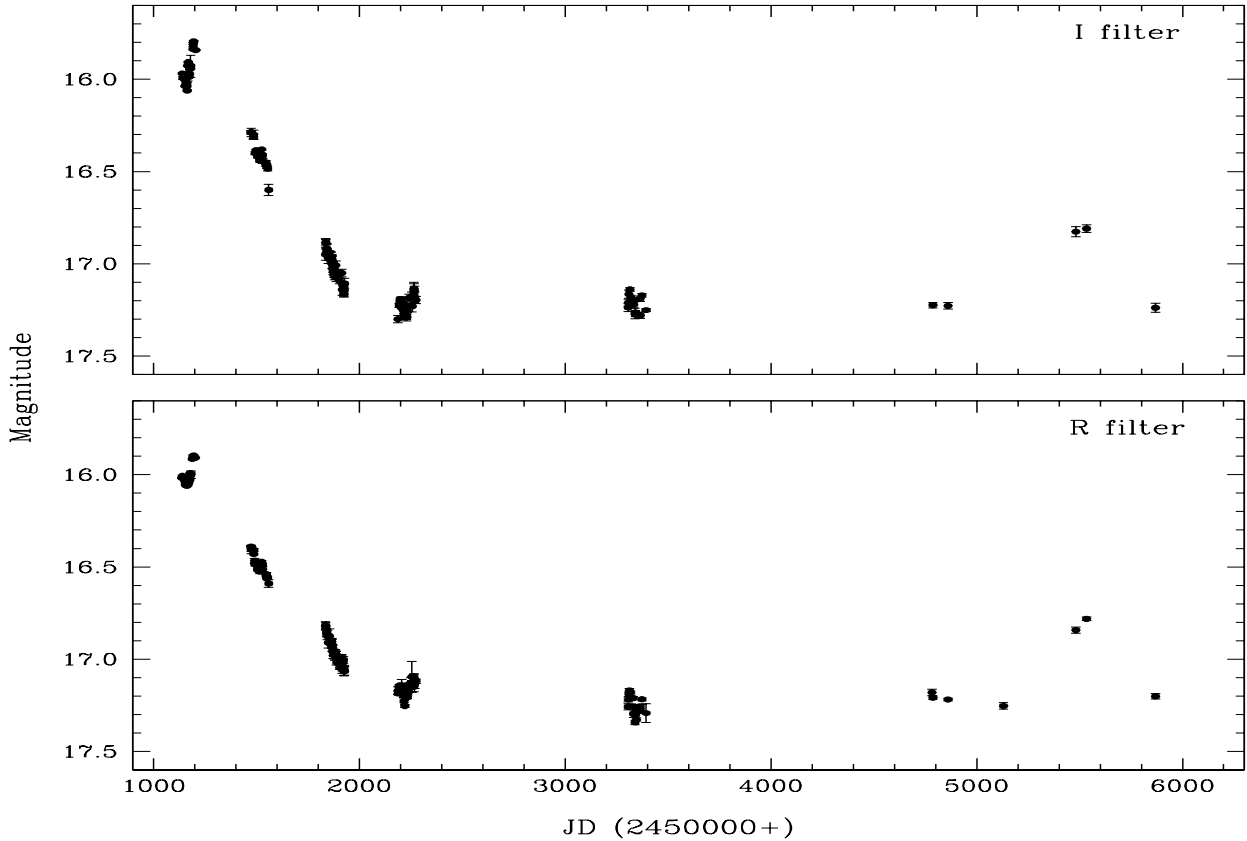}
\caption{The $R$ and $I$ photometric variation as a function of time during the observing period 1998 to 2011. Lower and upper panels represent the $R$ and $I$ bands, respectively.}
\label{phot_lc}
\end{figure*}
%
\subsection{Temporal evolution}\label{lc}
The light variations of {\sl AF And} in $R$ and $I$ bands are shown in Figure~\ref{phot_lc}. Here, the errors in data points are too small to be seen in comparison to the large amplitude of magnitude variation. From Figure~\ref{phot_lc}, it is quite evident that {\sl AF And} was in outburst phase in the first year of our survey itself when the star brightened by about 1.5 mag in comparison to the quiescence phase which we missed prior to the outburst but captured at the later stage. It is important to note that this star also went through an eruption phase during 1987-1991 when it also brightened by same magnitude \citep{Szeifert1996}. In the present data, the star was found to show maximum brightness in both the filters simultaneously on 16 January 1999 (at JD 2451195.1) having $R_{max}=15.896\pm0.003$ mag and $I_{max}=15.794\pm0.004$ mag. It is also noticed that apart from a normal fall in the stellar brightness, random fluctuations in brightness with amplitude of about 0.1 mag within a time scale of a few weeks are also present in the light curves. From the brightness variations illustrated in Figure~\ref{phot_lc}, one can observe that three different evolutionary phases are present on different time scales. In the following subsections, we discuss the photometric evolution of {\sl AF And} during these three phases.
\subsubsection{Primary outburst}
In the primary outburst phase of our observations, the flux increased by 0.23 mag in $R$ band and 0.27 mag in $I$ band in less than two months and reached to peak brightness. At this stage, the eruption triggers an enhanced mass outflow that leads to the formation of an extended atmosphere or "pseudo-photosphere" \citep{Szeifert1996}. The increasing flux during the outburst indicates that the star is approaching the instability. It is suggested that before the star reaches the Eddington limit, the luminosity is heavily dependent on mass as $L \propto M^3$, but it changes to $L \propto M$ as the star approaches Eddington limit \citep{Owocki2008}. Furthermore, the radiation pressure dominates gas pressure as the star approaches Eddington Limit and mass loss takes place. During the decay phase post peak brightness, brightness of the star decreases gradually and it took almost 3 years to reach the quiescent phase. Although there are gaps during the decline phase due to our irregular observations in the survey, it is quite evident that we indeed caught the star during the entire decay phase as well as when it finally settles on the quiescent phase. During this period, the star reduced in brightness by about 1.5 mag in both $R$ and $I$ filters. Over a duration of 3 years, the visual brightness difference between the two extremes is found to be about 1.5 mag in our photometric data for the star {\sl AF And}. 

In the initial stage, it is believed that the star throws out significant amount of stellar material in the stellar atmosphere resulting in a high mass loss, however, it slows down gradually in the post-outburst phase. Here, the star expands slowly with a speed of around 100-200~km/s making the star cooler which resembles a supergiant. The wind velocity primarily depends on the size of the stellar photosphere which dictates the visual brightness of the star. The optical density is inversely proportional to $T_{\textrm{eff}}$ \citep{deKoter1996}. Hence the temperature of the star starts to rise during the post-outburst phase as a consequence of increase in the stellar size.
\subsubsection{Quiescent phase}
The brightness decline phase of {\sl AF And} lasted for about 3 years and by the end of year 2001, the star has returned to its quiescent state. During this phase, brightness of {\sl AF And} remains almost constant though small short-term variability ($<$0.1 mag) have been noticed. Although we do not have enough phase coverage during the quiescent phase due to end of our microlensing survey program but from the intermittent follow-up observations taken over a period of about 3000 days, we do not see any significant change in brightness in both the filters. During this phase, even if the winds are slow, star is hotter than a normal hot supergiant with similar spectral type. In its effect, the star sheds some of the mass which can go up to $10~M_\odot$ \citep{Owocki2008} and thus the effective gravity of the star decreases that causes escape velocity of the star to decrease \citep{Humphreys2014b}. At visual minimum, the temperature of the star reaches $>$30,000 \,K as we have seen in the present case of {\sl AF And}. This stage may last for several years with flux remaining fairly constant and high temperature as described above.
%
\begin{figure*}
\includegraphics[width=17.0cm,height=7.5cm]{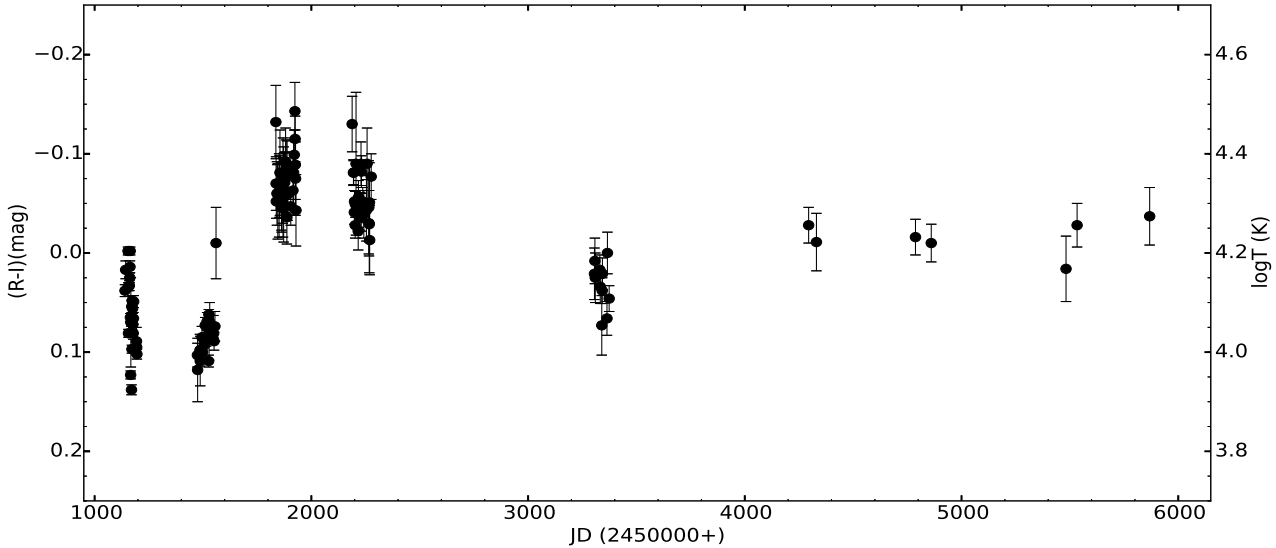}
\caption{The colour and temperature variations as a function of time during the entire monitoring period.}
\label{col}
\end{figure*}
%
\subsubsection{Secondary outburst}
After years of quiescent phase, {\sl AF And} again showed secondary outburst on 12 October 2010 in our data (at JD 2455533.2) when its brightness increased by about 0.44 mag in both $R$ and $I$ bands in comparison of its quiescence phase. Such short amplitude variation lasting over a few months are not unusual in LBVs \citep{Clark2012,Humphreys2014a}. Unfortunately, we had no observations prior to 12 October 2010. The next observations could be done only on 02 December 2010. On both the observed nights, the star has shown enhanced brightness. We collected 5 frames each in $R$ and $I$ bands on two observing nights during the secondary outburst phase. The star has shown brightening in all the 10 frames. In the light curves presented by \citet{Lee2014} using the PANSTARS data taken during mid-2010 to 2012, {\sl AF And} had indeed shown a brightness variation of about 0.6 mag in both PS1 $r$ and $i$ bands where maximum brightness was attained around mid-December 2010 before fading down to the quiescence phase about one year later. This is in confirmation of our observations taken during the same period. 
\subsection{Colour Excess}\label{reddening}
In order to de-redden the apparent magnitude for extinction, we need information of reddening in the direction of our observed field. Using the $R$ and $I$ bands $P-L$ diagrams of classical Cepheids found in our target field of M31, we determined the reddening value as $E(R-I)=0.16\pm0.11$ mag \citep{Joshi2003}. Since reddening estimation in our previous study has large uncertainty due to significant scatter in the $P-L$ diagrams of the Cepheids, we did not consider this value in our present analysis. Instead reddening, $E(B-V)$, was estimated based on the extinction maps provided by \citet{Schlegel1998} and \citet{Schlafly2011} using Galactic Dust Reddening and Extinction computing facility on NASA/IPAC Infrared Science Archive \footnote{\url{https://irsa.ipac.caltech.edu/applications/DUST/}}. The resulting reddening values are given in Table~\ref{tbl:basic_parameters}. Using the values listed in the table, we adopted an average reddening of $E(B-V)=0.36$ mag in our analysis which contains both foreground reddening due to the Milky Way as well as reddening due to the disk of M31. The corresponding extinction was obtained using the relation $A_V = R\times E(B-V)$ where $R$ was taken as 3.07 \citep{Fitzpatrick1999}. We obtained an extinction of $A_V=1.11$ mag in the direction of our observed field which is significantly large but considering the location of {\sl AF And} in the spiral arm of M31, it is well expected.
\subsection{Colour and temperature evolution}\label{cev}
Amongst the total of 169 nights observed in the present study, {\sl AF And} was monitored in both $R$ and $I$ bands on 126 nights. For these nights, we determined $(R-I)$ colour of the star. The variation in the colour during our observations is illustrated in Figure~\ref{col} where we can see a change in colour from 0.14 mag in outburst phase to -0.14 mag in the quiescent phase. Although about 0.28 mag variation is seen in the colour of the star during its evolutionary phase in our data, however, most of the variation in colour has been observed in the first 3 years of our survey. One can notice that the colour evolution does not exactly follow the light curve variation. It is suspected that dust formation during the outburst may also have a strong influence on the colour evolution. To get a better physical interpretation of the evolution of {\sl AF And}, photometric colour $(R-I)$ was converted to the corresponding effective temperature ($T_{eff}$). For this, first we converted $(R-I)$ into $(B-V)$ colour using the relation $(B-V) = (R-I)/0.55$, assuming a normal reddening law. To convert $(B-V)$ to $T_{eff}$, we derived the following colour-temperature relation for the supergiants using the Table 4 of \citet{Flower1996}.
\begin{equation}
log~T_{eff} = 4.01258-1.05432~x+2.13061~x^2-2.4917~x^3
\label{eq1}
\end{equation}
where $x$ is the colour $(B-V)$ of the star. In Figure~\ref{col}, we also illustrate the corresponding temperature variation of the star as a function of time on the right side of y-axis. The resulting temperature of the star only gives an approximate value but good enough to analyse general properties of {\sl AF And} which are described below for three different phases of its evolution. 
\subsubsection{Primary outburst}
During the rising phase of {\sl AF And}, colour of the star gradually increases and reached a maximum value of $(R-I) = 0.14$ mag which corresponds to a temperature of about 7050\,K. Here the temperature of {\sl AF And} is similar to that of the F spectral type stars. According to \citet{Humphreys1994} and \citet{deKoter1996}, LBVs attain temperature of around 7000\,-\,9000\,K at the time of peak brightness during the outburst phase when star reaches to cold state. It is evident that the maximum positive colour which represents coolest temperature exactly coincides with the time of peak brightness of the star. As the eruption subsides, an inversion in the colour was found when $(R-I)$ changes from maximum positive value of 0.14 mag to maximum negative value of -0.14 mag in about 500 days indicating that the temperature in the photospheric region was increasing.

During the post-outburst phase, the mass loss rate increases and due to higher mass loss rate and lower wind velocity, the winds get optically thin \citep{Garcia1997}. It causes temperature to rise gradually \citep{Humphreys1994}. In the case of {\sl AF And}, it reaches up to $T_{\textrm{max}} >$ 30,000\,K as seen in the present photometric observations. Since, the bolometric luminosity remains fairly constant in LBVs \citep{Wolf1989,Humphreys1994}, temperature rise is supplemented by decrease in the radius which is caused by the high mass loss due to stellar winds.
\subsubsection{Quiescent Phase}
In the quiescent phase, {\sl AF And} settled with almost constant magnitude and effective temperature of $>$30,000\,K where it resembles a spectral type of O-type supergiant. Here, the outer envelope of the star which has acquired high temperature due to mass loss continues to be in a minimum state which does not cause much change in the luminosity and colour as seen in Figures~\ref{phot_lc} and ~\ref{col}, respectively. The star may remain for an extremely long period in this state. After long quiescent phase of {\sl AF And}, the wind again starts to radiate more i.e. become less opaque due to reduced mass loss rate \citep{Garcia1997} hence brightness may increase.
\subsubsection{Secondary Outburst}
We caught a secondary outburst when stellar brightness increased by about 0.44 mag in late 2010. During this phase, the colour of {\sl AF And} was found to be $0.02\pm0.03$ mag and $-0.03\pm0.02$ mag on the two observed nights in our observations separated by almost 2 months. The total colour variation during this phase is only $0.05\pm0.04$ mag which corresponds to a change in temperature of about 2200\,K. However, due to poor sampling of our data, this does not reflect a total change of temperature during the secondary outburst phase. In a better coverage of this outburst in the PANSTARS data collected between 2010-2012 \citep{Lee2014}, {\sl AF And} has shown a PS1 $(r-i)$ colour variation of 0.25 mag at the time of maximum brightness.

In Table~\ref{tbl:mag_col_vari}, we summarize the amplitudes of variability in $R$ and $I$ bands as well as in $(R-I)$ colour during various phases of evolution of {\sl AF And} in our data.
\input{table04.tex}
\subsection{Colour-Magnitude variation}\label{cmd}
\begin{figure}[t]
\centering
\includegraphics[width=9.0cm, height=11.0cm]{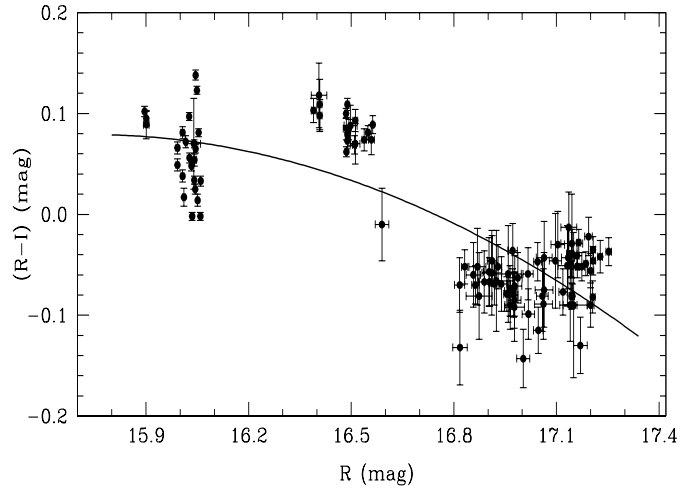}
\vspace{-4.7cm}
\caption{The variation in $(R-I)$ colour with the brightness in the $R$ band. The errorbars in both colour and magnitude are shown for each point. The polynomial fit is also drawn by a continuous line.}
\label{col_mag}
\end{figure}

The $(R-I)$ colour variation of {\sl AF And} during the primary outburst phase in our observations with respect to corresponding $R$ band magnitude is shown in Figure~\ref{col_mag}. It is apparent that despite the irregular colour variation, there is a trend in the colour-magnitude diagram during the evolution of star from the outburst phase to quiescent phase. Although there are large uncertainties in the colour determination when star is fainter but, in general, we find that {\sl AF And} exhibits a redder-when-brighter (RWB) trend in its colour-magnitude diagram. This RWB behaviour is typical of a LBV characteristic as they become cooler during the outburst phase. In the early part of the outburst phase, the colour remains almost constant as can be seen in Figure~\ref{col_mag}. The temperature of the LBV during the quiescent phase reaches above 30,000\,K making it bluer. 
%
It is noticed that the colour-magnitude diagram becomes steeper for fainter magnitudes. In the figure, the variation in the $(R-I)$ colour with respect to $R$ magnitude may be represented by a quadratic polynomial equation of the following type
\begin{equation}
(R-I) = -0.08~(R-15)^2 + 0.11~(R-15) + 0.04
\end{equation}
$$
~~~~~~~~~~\pm0.03~~~~~~~~~~~~~~\pm0.10~~~~~~~~~~~~\pm0.07
$$
\noindent A best fit shown by a continuous line in Figure~\ref{col_mag} yields a $\chi^2$/ndf of 0.04. A similar trend was also shown by \citet{Sholukhova2011} and \citet{Solovyeva2019} in their analysis of LBVs in the galaxies M33 and NGC~4736, respectively. Our study further ascertains the characteristic RWB nature during the LBV phase that suggests the colour of the LBV star becomes bluer as a consequence of increase in the temperature of stellar photosphere in its quiescence phase.
\subsection{Bolometric magnitude}\label{bm}
One of the important parameter to evaluate the extent of outburst of any eruptive variable is its peak bolometric magnitude. In the case of LBV outburst, this information is critical for evaluating the extent to which the star exceeded its own Eddington limit though temporarily with small adjustments in opacity or stellar structure \citep{Smith2011}. Using the extinction value of $A_V$ = 1.11 mag as determined earlier, we estimated an extinction of $A_R$ = 0.95 mag and $A_I$ = 0.71 mag assuming the normal extinction law in the direction of our observed field. Considering a distance of 750~kpc for M31 \citep{Freedman2001}, we estimated a maximum visual magnitude of $M_R \approx -9.43$ mag and $M_I \approx -9.30$ mag which correspond to a maximum visual magnitude of $M_V \approx -9.49$ mag. Since bolometric correction at peak brightness is negligible \citep{Smith2011}, the above maximum absolute magnitudes correspond to $M_{bol} \approx -9.49$ mag which yields a luminosity of $log(L/L_{\odot}) \approx 5.70$. The peak luminosity and colour during the outburst phase suggests a F-type spectrum for {\sl AF And}. \citet{Szeifert1996} derived a luminosity of $log(L/L_{\odot})$ = 5.9 to 6.1 for {\sl AF And} which is consistent with our estimated value. As the brightness variation in the decline phase is compensated by the change in bolometric correction, the total luminosity of the star remains almost constant \citep{Wolf1989, Humphreys1994}. Hence bolometric magnitude during the quiescent phase also remains at the same level of about -9.49 mag. Here, we caution readers that the largest uncertainty in the estimation of bolometric magnitude or total luminosity comes through the extinction derived in the direction of the target field.
\subsection{Rate of decline}\label{rd}
As the rise time in eruptive variables is poorly constrained most of the time due to sudden outburst, but their temporal evolution during the fading stage is relatively better characterized. The photometric light curve illustrated in Figure~\ref{phot_lc} shows a continuous decrease in brightness from the maxima before it becomes steady. The rate of decline is generally quoted as either a fall of 2 mag ($t_2$) or 1.5 mag ($t_{1.5}$) from its peak brightness. In the case of {\sl AF And} the total fall in brightness is $\sim$ 1.5 mag between the peak brightness phase and quiescent level hence we determined $t_{1.5}$ for this star. It is found that the maximum and minimum brightness respectively fall on $JD$ = 2451195 and 2452222 in our data. This means that it takes approximately 1027 days between the time of maximum brightness and the time when the star fades back to its quiescent level. Using a linear approximation, we determined the decline rates of {\sl AF And} in $R$ and $I$ bands during its outburst between 1998 to 2001 to be:
\begin{equation}
t_{1.5} (R,I) \sim 0.0015 ~ \tt{mag~day}^{-1}
\label{eq5}
\end{equation}
Although we did not have a complete and uniform coverage during the post-outburst phase of {\sl AF And}, the estimated decline rate gives an idea how slow the star fades from the peak brightness in its LBV phase. It is important to note here that decay rate does not exactly correlate with the amount of total mass ejected from the LBVs \citep{Smith2011}. Recent estimate of mass loss in two well known LBVs in the Milky Way $\eta$ Carina and P Cygni are about 10 $M_\odot$ and 0.1 $M_\odot$ respectively despite both having similar duration of about 10 years for their major eruptions \citep{Smith2003,Smith2006}. Although it is well accepted that the eruptive nature of LBVs is the consequence of their approaching or exceeding the classical Eddington limit \citep{Smith2011}, it is still not very clear which underlying physical mechanism drives the extent of mass loss during the major eruptions of these stars.
%
%
\begin{figure*}
\centering
\includegraphics[width=17.0cm, height=9.0cm]{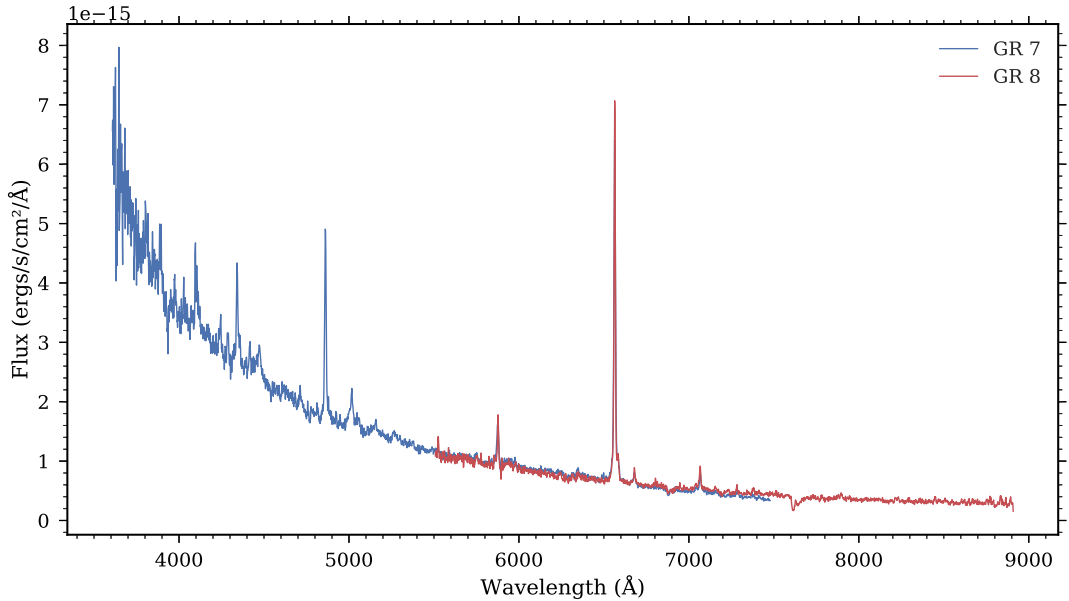}
\caption{Flux calibrated spectrum of AF And in the wavelength range 3800-8400~\r{A}. The observations with grisms gr7 (blue solid line) and gr8 (red solid line) were taken at mJD equal to 2457642.29087 and 2457642.32407, respectively when the star was slightly brighter than its quiescent phase \citep{Martin2017a}.}
\label{fig:flux_calibrated}
\end{figure*}

\begin{figure*}
\centering
\includegraphics[width=17.0cm, height=7.0cm]{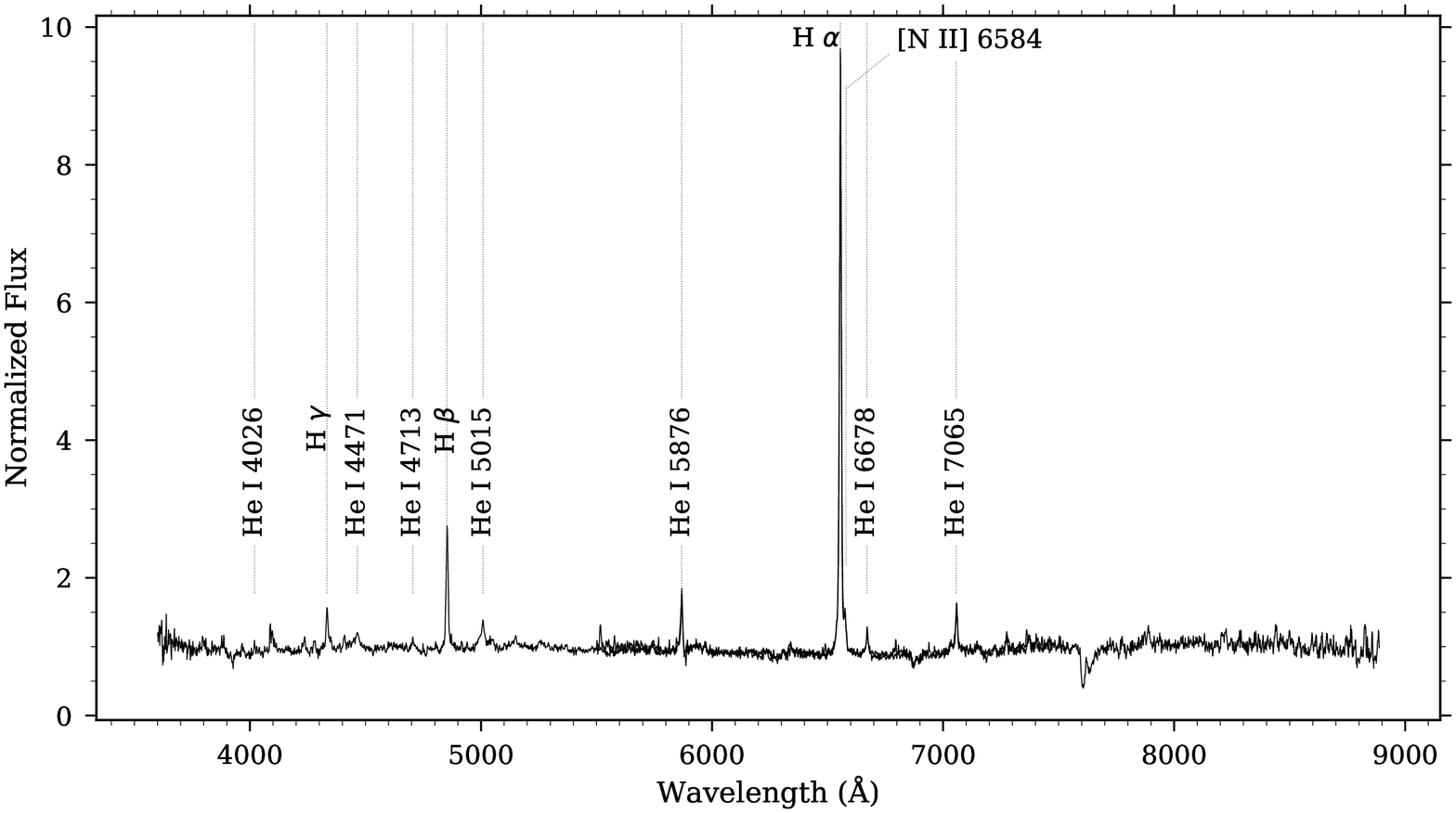}
\includegraphics[width=17.0cm, height=7.0cm]{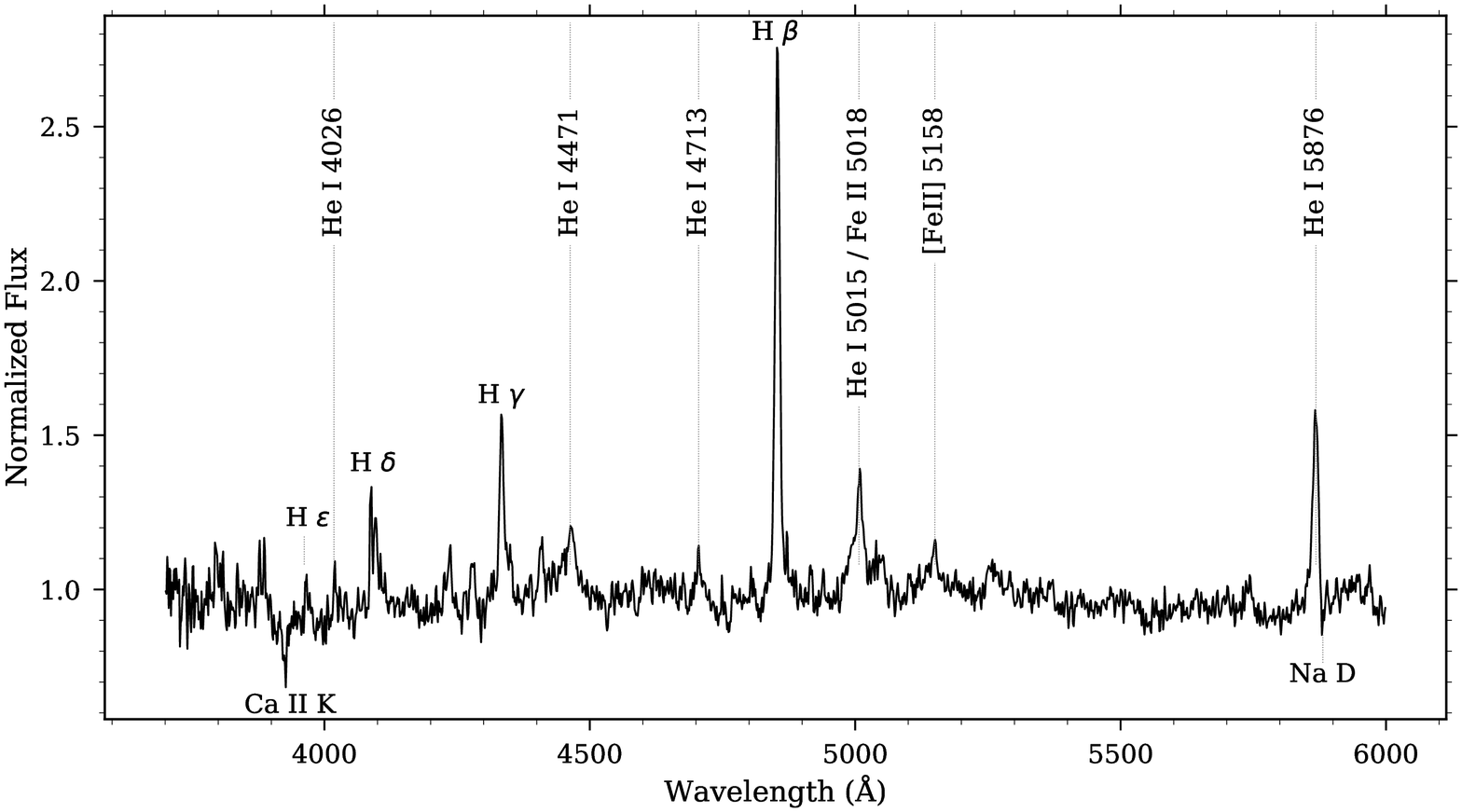}
\caption{Continuum-normalized spectrum of {\sl AF And} with various prominent spectral features marked with a dashed line. In the top panel the spectrum is plotted in the whole wavelength coverage of our observations. For clarity, we show the spectrum in the range 3800-6000 $\r{A}$ in the bottom panel to highlight the strong emission lines.}
\label{fig:flux_normalized}
\end{figure*}
%
\section{Spectroscopic Data}\label{spec}
Spectroscopic studies of {\sl AF And} in the past also reveal its spectroscopic variability in quiescent phase. \citet{Gallagher1981} note the presence of more FeII and [FeII] lines as compared to the previous observations by \citet{Humphreys1975,Humphreys1978}. To investigate this aspect, we carried out spectroscopic observations of {\sl AF And} using Hanle Faint Object Spectrograph Camera (HFOSC) mounted on 2.0-m Himalayan Chandra Telescope (HCT), operated by Indian Institute of Astrophysics (IIA) at Hanle, Ladakh, India in the low-resolution mode on the night of 10 September 2016. We used SITe CCD of 2K$\times$4K pixels with pixel size of 15$\mu\times$15$\mu$ and image scale of 0\arcsec.29/pixel. We used grism gr7 (3800-6800~\r{A}) and slit width of 0$^{\prime\prime}$.77 (67 $\mu$), giving a spectral resolution of R$\sim$1300. Far visible region of the spectrum was collected using gr8 (5800\,-\,8400~\r{A}) having a slit width of 1$^{\prime\prime}$.15 (100 $\mu$) and a resolution of $R \sim 2200$. Our optical spectra of {\sl AF And} thus cover the wavelength range from 3800 to 8400~\r{A} although there was significant overlap from 5800 to 6800~\r{A} between two set of grisms. The bias, flat and arc spectra were observed on the same night for the calibration purposes. Fe-Ne spectra for the wavelength calibration in gr7 could not be taken on the observation night because of some technical issues. Therefore lamp spectra were taken from the previous nights for gr7 wavelength calibration. Feige110 was observed as a standard star in both grisms for the flux calibration on the same night. All steps for the spectroscopic data reduction were performed under IRAF environment. We also checked the spectra for the contamination problem but did not find any nearby objects within 10 arcsec of our target star. This rules out any contamination of light in our spectral observations. Final flux-calibrated spectra and dereddened gr7 and gr8 spectra are shown in Figure~\ref{fig:flux_calibrated}. The present spectra corresponds to the phase when star was going towards the outburst phase from the quiescent phase as can be seen from \citet{Martin2017b}.
%
\begin{figure*}
\centering
\includegraphics[width=17.0cm, height=8.5cm]{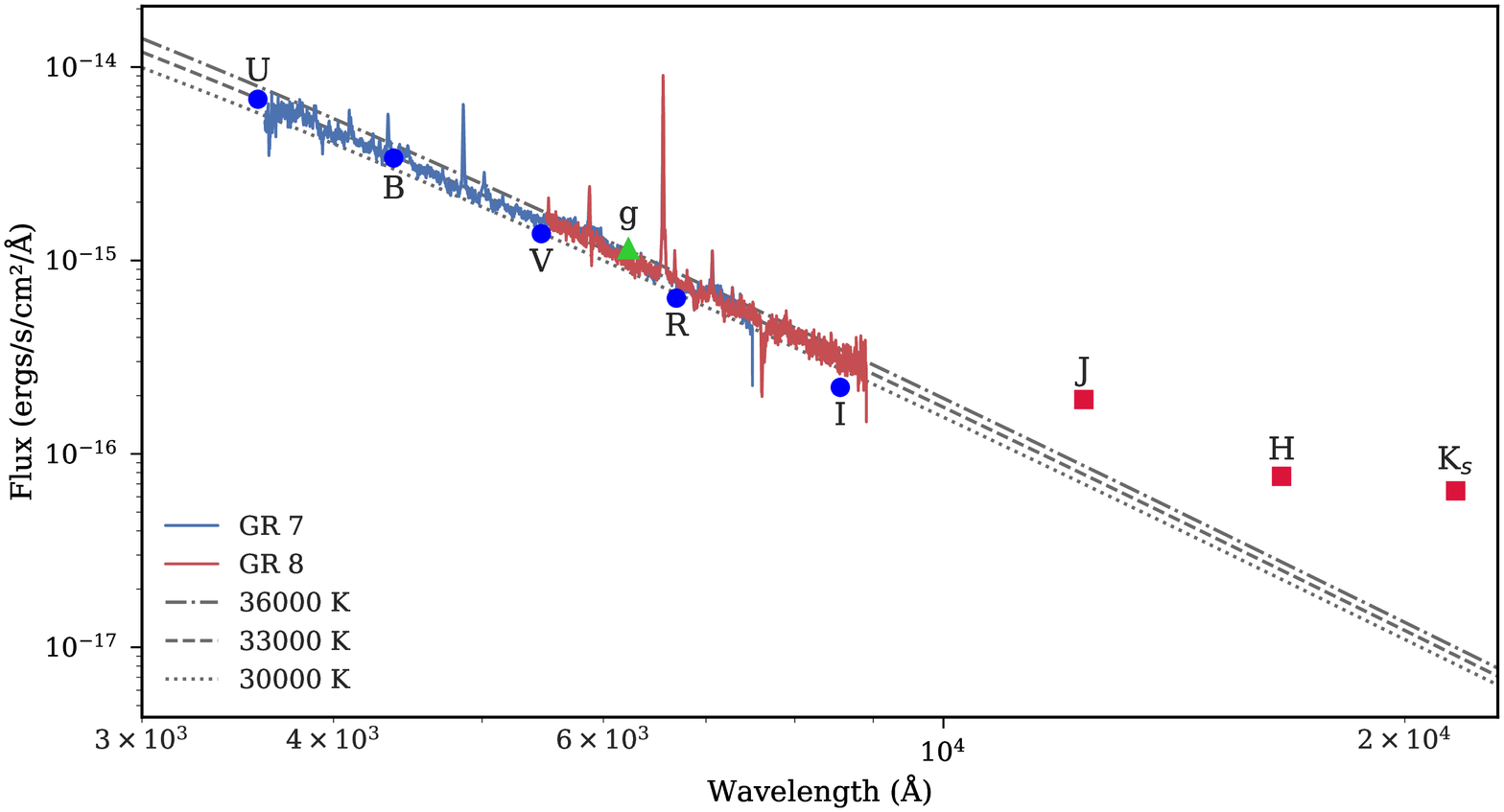}
\caption{Spectral Energy Distribution of {\sl AF And} from 3000 \r{A} to 2.2 $\mu$. Solid lines represent the flux-calibrated and rescaled gr7 (blue) and gr8 (red) spectra whereas other line styles in grey color show the Planck curve for different temperatures as indicated in the legend. Flux values obtained from reddening corrected $UBVRI$, GAIA $g$ band and 2MASS $JHK_s$ photometry are represented by blue circles, green triangle and red squares respectively. Errors corresponding to flux values are very small ($\sim 10^{-17}$) and therefore not shown in the figure.}
\label{fig:final_fit}
\end{figure*}
%
\section{Spectroscopic Analysis of AF And}\label{spec_ana}
The LBVs not only reveal the photometric variability but also show different spectra that depends upon the evolutionary state they are in. In the outburst phase, their spectrum resembles a F to A-type supergiant whereas, in the quiescent phase, they show the spectrum similar to a hot supergiant with H, He and FeII, prominently seen in emission along with P-Cygni profile. Our observations were taken close to the quiescent phase and are marked with the presence of strong H Balmer (6563, 4861, 4340, 4101 \r{A}) emissions with very weak [N II] feature at 6584 \r{A}. For an easier line-identification, the continuum normalized spectra are shown in Figure~\ref{fig:flux_normalized} which highlights the presence of He I at $\lambda$4026, 4471, 4686, 5015, 5876, 6678 and 7065 \r{A} apart from other weaker emissions of He I. The spectrum is dominated by many FeII and [FeII] emission lines, a typical signature of LBV spectrum in the quiescent phase, but the prominent ones are observed at FeII $\lambda$5018 and [FeII] $\lambda$5158. A weak Ca II $\lambda$3933 absorption and Na D line are also present in the observed spectrum. These are shown in the lower panel of Figure~\ref{fig:flux_normalized}. Here, outburst and cooler phase of LBV spectrum is marked by the presence of Mg II $\lambda$4481 and other A-type absorption features which are missing from our spectra. This indicates towards the hotter surface of the star. The features in our spectrum match with the previous observations of \citet[their Figure~5]{Humphreys2014a} where He \textsc{I} lines with weak emissions of N \textsc{III} are observed. We also note the asymmetric line profile in the strong emissions, e.g. He I, and split in the Balmer lines which indicates the mass loss, winds, and outflow, a characteristic phenomenon of LBVs.
\subsection{Energy distributions and photospheric temperature}\label{spec_temp}
For estimating the photospheric temperature by fitting a black body spectrum on the spectral energy distributions (SED), we used the photometric data from \citet{Massey2006}. To correct the $UBVRI$ magnitudes for the interstellar extinction, color excess $E(B-V)$ is taken to be equal to 0.36 mag as mentioned in Section~\ref{reddening}. The same value is also adopted for the reddening correction of the observed spectra. Flux values corresponding to the photometric observations are shown in Figure~\ref{fig:final_fit}. The errors on photometric flux values are of the magnitude similar to the size of the symbols and hence have not been represented on the diagram. The calculated photometric flux values and optical spectra are fitted against the Blackbody curve for various values of temperature. After visual inspection of the SED fitting, we assign 30,000\,-\,36,000\,K as the estimated range of photospheric temperature for {\sl AF And}. The presence of infrared excess is also quite evident which is generally observed in the LBVs because of free-free emission and/or thermal emission from the dust in the NIR region \citep{Kraus2014,Humphreys2014a,Sholukhova2015}.

For an independent confirmation of our $T_{\textrm{eff}}$ estimate, we also try an online utility called Virtual Observatory SED Analyzer \citep[VOSA;][]{Bayo2008} hosted by the Spanish Virtual Observatory. It allows the user to input the photometric magnitudes in different bands as well as queries the existing photometric databases to retrieve the photometry for the object of interest. After constructing the SEDs, two independent approaches based on theoretical and empirical stellar models are explored for estimating $T_{eff}$. In the first approach, $\chi^2-$fitting is performed against the theoretical models from Kurucz model atmospheres \citep{Castelli1997} for different atmospheric parameters. Best fitted $T_{\textrm{eff}}\,=\,33,000\pm3000$\,K is returned corresponding to the minimum value of $\chi^2$. For the other method, empirical templates of flux calibrated stellar spectra, with spectral classes ranging from O5 to M9 (including 5 spectral templates of L type stars) at $R\sim2000$ and wavelength coverage of $3650\,-\,10200$ \r{A}, are taken from \citet{Kesseli2017}. These templates were created by the co-addition of individual stellar spectrum belonging to the same spectral class from Sloan Digital Sky Survey \citep[SDSS,][]{York2000}. Fitting the derived SED against Kesseli templates returns O7 spectral type as the best matching flux distribution, which is also consistent with the photometric temperature obtained in Section~\ref{cev}.
\subsection{Host galaxy metallicity}\label{Metallicity}
Several diagnostics exist for the metallicity measurements \citep{McGaugh1991,Kewley2002,Pettini2004,Pilyugin2005} which depend mostly upon the emission line ratios and calibration measurements. Systematic differences are obtained by using different calibration measurements \citep{Ellison2005}. \cite{Sanders2012} used a number of H II regions and planetary nebulae in M31 to provide metallicity estimates. They show that the ISM of M31 is mostly clumpy and inhomogeneous with large gradient in oxygen and [NII] abundances.  \cite{Sanders2012} metallicity estimates varied from 7.3 dex to 9.1 dex. Since the sensitivity of HFOSC is low towards the blue end, we were unable to detect [O III] 3727 \r{A} and 3729 \r{A} line. Furthermore, the [S II] 6717, 6731 \r{A} line are also not detectable. However, weak [N II] $\lambda$6584 \r{A} feature and [O III] $\lambda$5007 \r{A} line is resolved. So, we were able to measure the O3N2 index and N2 index for metallicity estimates. For the N2 index, \cite{Pettini2004} relates the metallicity by the following equation:
\begin{equation}
12 + log(O/H) = 8.90 + 0.57 \times N2
\label{eq6}
\end{equation}
where N2 = log([NII] 6584/H$\alpha$) and 95\% of the measurements being within $\pm$0.41 of the values as derived from the above equation. Using above equation, the metallicity value is estimated to be 8.89$\pm$0.41 dex for M31. \cite{Pettini2004} also gave the metallicity relation for O3N2 index given as:
\begin{equation}
12 + log(O/H) = 8.73 - 0.32 \times O3N2
\label{eq7}
\end{equation}
where O3N2 = log{([O III] 5007/H$\beta$)/([NII] 6584/H$\alpha$)} and measurements are within 0.25 of the best fit line. The estimated metallicity value of M31 is 8.73$\pm$0.25 dex using above method. \citet{Stasiska1998} estimated a metallicity value of 8.64$\pm$0.23 dex for the M31 bulge which are within the error bars of our estimated value.
\subsection{Terminal velocity}\label{ter-vel}
Wind speed or terminal velocity is often calculated from the blue edge of the P Cygni profile \citep{Humphreys2014a}. We also see a weak P-cygni feature in the HeI emission line of {\sl AF And} in Figure~\ref{fig:flux_calibrated}. A dip is also seen in the right side of the P-cygni profile of {\sl AF And} which is possibly originating due to the blend of Na D doublet. From the left side absorption minima of HeI 5876 \r{A} line, we find the wavelength corresponding to the central dip of {\sl AF And}. Since the rest wavelength is already known, applying the Doppler shift formula, we estimate a terminal velocity of $300\pm35$~km~sec$^{-1}$. We also performed a Gauss fit over the blue-edge of {\sl AF And} and the terminal velocity is obtained to be $280\pm30$ km~sec$^{-1}$. Previous studies on {\sl AF And} gave an estimated terminal velocities between 100 - 300 km~sec$^{-1}$ measured from blue-edge of HeI 5876 line \citep{Leitherer1994,Stahl2001,Groh2009}. Some of the galactic LBVs, AG-Car had shown low velocities during the quiescent phases \citep{Smith1994}. \citet{Groh2009} found a very low velocity of 105 km~sec$^{-1}$ for {\sl AF And} during its 2001 minima. In quiescent phase of {\sl AF And}, \citet{Szeifert1996} has estimated a terminal velocity of 150 km/s. The value obtained by us is slightly higher than their value as we have taken spectroscopic observations when star was in the rising branch of the outburst phase. However, the values obtained by us and \citet{Groh2009} are much less than the typical value for WNL and late-WN stars \citep{Crowther1995a, Crowther1995b} which varies between 400 to 1300 km~sec$^{-1}$ and also between 400 - 1000 km~sec$^{-1}$ for early type supergiants \citep{Mokiem2007}. The smaller terminal velocity estimated in the present study is indicative of dense and slower winds around LBVs. Since they have lost most of their mass during S-Dor type variability phase and in giant eruptions, their terminal velocities are lower than most supergiants and late WNLs.
\subsection{Mass loss rate}\label{mass-loss}
We estimate the mass-loss rate of the luminous blue variable star assuming the luminosity of the ejecta-CSM interaction is fed by energy imparted at the shock front. The progenitor mass loss rate $\dot{M}$ can be broadly calculated using the relation of \cite{Chugai1994}:
\begin{equation}
\dot{M}=\frac{2L}{\epsilon}\frac{v_w}{v_{LBV}^{3}}
\label{eq8}
\end{equation}
where $\epsilon < 1$ is the efficiency of converting the shock's kinetic energy into visual light (an uncertain quantity), v$_w$ is the velocity of the pre-explosion stellar  wind, v$_{LBV}$ is the velocity of the shell undergoing eruptions and L is the bolometric luminosity of {\sl AF And} in the same phase. As mentioned in section \ref{bm}, we estimated a bolometric luminosity value of log$_{10}$L = 5.7 L$_{\odot}$ which essentially remains constant during both the quiescent and eruptive phases. We also estimate a typical wind velocity of v$_{w} \approx 280$ km~sec$^{-1}$ which is also taken into account for the estimations. The H$\alpha$ component has a full width at half maxima of 54.165 \r{A} which corresponds to a velocity of 2476 km~sec$^{-1}$. The efficiency factor $\epsilon$ is taken to be 0.5 (at lower optical depth the efficiency may be less; see \citet{Smith2007}). Using above values, the estimated mass loss rate in this phase is found to be about 2.2$\times$10$^{-4}$ M$_{\odot}$ yr$^{-1}$. Many authors have reported slightly smaller mass-loss rate than our estimate \citep{Humphreys1994,Szeifert1996} but most of them have estimated mass-loss rate during the quiescent phase when mass-loss is found to be smaller than the outburst phase.

We summarize stellar parameters obtained in the present study in Table~\ref{tbl:param}. Since many uncertainties are involved in the determination of these parameters, our values can only give an approximate estimates of given parameters. Nevertheless, they show a broad characteristics of the LBVs found in the M31 and M33.
\input{table05.tex}

\section{Discussion}\label{discuss}
The LBV star {\sl AF And} in M31 has been known to go through major eruptive phase quite often since a long time and one of the most studied object in the past \citep{Gallagher1981,Sharov1990,Szeifert1996,Humphreys2014b,Humphreys2017}. Even though significant variability on smaller scale of a few years to longer scale of decades have been noticed but no clear periodicity was found in the light curve variations of this star hence it is very difficult to anticipate onset of its eruption. However, thanks to our microlensing survey program towards the M31 disk, where our target field happens to fall on the location of {\sl AF And}, we serendipitously caught this star going through the outburst phase in 1998 when it has shown a major brightness change of about 1.5 mag in a time span of about 3 years. {\sl AF And} has shown an absolute maximum (cold state) of visual brightness during early 1999 and absolute minimum (hot state) in late 2001. During the same period, its $(R-I)$ colour has changed by about 0.28 mag when the temperature has increased from about 7000\,K at the time of peak brightness to over 30,000\,K during the quiescence phase. Our sparse follow-up monitoring of {\sl AF And} after the end of our microlensing survey program has shown a secondary outburst in this star in late 2010 followed by a quiescent phase again. During this time, an increase in brightness of about 0.44 mag was noticed in our photometric data. This star has also been undergoing many eruptions in the past as reported in many previous studies where major outbursts were normally observed with an amplitude of 1 to 2 mag while minor outbursts were observed with 0.4 to 0.6 mag photometric variations \citep{Humphreys1994,Szeifert1996,Kurtev1999,Lee2014}.

The evolutionary outburst phase has been characterized by the huge mass-loss rate due to stellar wind resulting in the stellar photosphere. During this phase, the mass loss causes the wind to become optically thick and photospheric temperature goes down. The pseudo-photosphere is formed around them, which is swept by the strong and fast winds. Although, star brightened by about 1.5 mag during the outburst phase in comparison of quiescence phase but it is believed that the total luminosity of LBV does not change much between the two phases. Even though the photometric coverage during the quiescent phase is quite patchy, it is found that {\sl AF And} resembles with a typical O-type supergiant where photospheric temperature has gone beyond 30,000\,K. Estimating an average interstellar extinction of $A_V=1.1$ mag, we obtained a maximum visual magnitude of $M_V \approx -9.49$ mag corresponding to a maximum luminosity of $log(L/L_\odot) = 5.7$ assuming no bolometric correction at the peak. 

Comparing the observed spectrum with respect to the one presented by \citet{Humphreys2014a} reveals its spectroscopic evolution. \citet{Humphreys2014a} spectrum shows no signature of FeII and [FeII] emissions whereas in our spectrum we see two strong Fe emissions. FeII $\lambda$ 5018 is blended with He I $\lambda$5015 and [FeII] at 5158 \r{A} is clearly distinguishable. Weaker FeII and [FeII] emissions are also present in the wavelength range 4900-5400 \r{A}. We observed a weak P-cygni profile in the emission lines of our observed spectra as has also been found by \citet{Humphreys2014a}. P-cygni profile with absorption at the blue end and emission at the red end suggests an outflow in the photosphere. For estimating the photospheric temperature using SED fitting, we used the $UBVRI$ magnitudes from \citet{Massey2006}, $JHK_s$ magnitudes from the 2MASS survey \citep{Skrutskie2006}, and $g$-band magnitude from the GAIA survey \citep{Gaiamission}. From the SED analysis, we determined a photospheric temperature of 33,000$\pm$3000\,K which happens to be close to the photometric temperature obtained through the colour information during the same phase. This is also the typical temperature most of the LBVs exhibit in their hot state.

The mass-loss takes places continuously in these stars through the incessant wind in their photosphere. Using the empirical mass-loss estimation, we found that {\sl AF And} has a high mass-loss rate of 2.2$\times10^{-4}$ M$_{\odot}$ yr$^{-1}$. Although mass-loss rate varies during the eruptive and quiescence states but it is dominant mode in the former one. In fact it is believed that these massive young stars have shed bulk of their initial masses in their supergiant phase hence their effective gravities and escape velocities are now much lower \citep{Humphreys1994, Humphreys2014a}. In recent evolutionary tracks considering mass loss and stellar rotation, \citet{Ekstrom2012} has shown that these stars would have shed about half of their initial masses by now and reached close to their Eddington limit. Since LBVs undergo significant mass loss during their evolution, they accumulate vast amount of material in their circumstellar environment ejected via wind. When a black body spectrum is placed in their energy distributions, most of the LBVs show infrared fluxes lying above the theoretical curve \citep{Kraus2014}. For the best fit of 33,000$\pm$3000\,K black body curve in the SED profile, {\sl AF And} also exhibits an infrared excess in its longer wavelength. The excess in LBVs is believed to be due to free-free emission and/or thermal emission from their stellar winds in the extended atmosphere \citep{Humphreys2014a,Sholukhova2015}. This in fact, along with their location in the spiral arms, make LBVs observations favourable in the longer wavelengths.

Unlike B-type supergiants and Of/WN stars, the LBV stars are well known to have low wind speeds during their quiescence phase where about 100-200 km s$^{-1}$ velocities were found \citep{Smith1994, Groh2009}. However, wind velocities are found to be relatively larger in the quiescent phase manifested in the form P-Cygni profile of emission lines. Though we do not observe as strong P-cygni profile in the emission lines as seen in the spectrum by \citet{Humphreys2014a}, but, an asymmetry towards the blue end exists in all prominent H and He lines. Based on our spectral analysis of a P Cygni feature in the HeI emission line, we estimated wind terminal velocity of about 280 to 300 km~s$^{-1}$ with an uncertainty of 30 to 35 km~s$^{-1}$. We note here that a more precise estimate of terminal velocity requires high-resolution spectroscopy comprising high S/N signal in the P Cygni profile but our low-resolution spectrum still provides an approximate estimation of the outflow speed which is well in agreement with the previous studies on such objects.
\section{Summary}\label{conclu}
The primary aim of the `Nainital Microlensing Survey' was to search for microlensing events in the direction of M31. However, the vast amount of data also enabled us to identify a substantial number of variable stars and optical transients in the target field of M31. Here, we carried out a long-term photometric analysis of the prominent LBV star {\sl AF And}. The study of LBVs is very important as only a small number of such stars are known making them unique and interesting objects. As we were fortunate enough, an outburst was caught in {\sl AF And} in the beginning of the survey itself. Our photometric analysis of the data taken for about 5000 days revealed that {\sl AF And} has gone through two outbursts, one major and other minor, in a time span of about 11 years. The main results of our study are summarized below.
\begin{enumerate}
\item {\sl AF And} has been caught in the outburst phase in 1998 when we started our survey to detect mircolensing events in the direction of M31. 
\item The star was in a fairly bright state at the start of our observations and reached peak brightness around mid-January 1999 before fading down to the quiescent phase over the period of 3 years.
\item We found a magnitude variation of about 1.5 mag between its primary maxima and minima and a colour variation of about 0.28 mag during this period.
\item The colour and temperature analysis of the star during eruption phase has shown that star was cooler at the time of peak brightness phase when its temperature decreased to about 7000\,K. During the quiescent state, the temperature rose to beyond 30,000\,K which is the typical characteristic temperature of such stars. For the {\sl AF And}, we also estimated $T_{eff}$ from SED which matches well with its photometric temperature.
\item The star exhibits a very slow decline of about 0.0015 mag~day$^{-1}$ before settling in the quiescent phase where it did not show any significant brightness variation over the next 9 years.
\item Our follow-up monitoring of this star has shown a secondary outburst in the year 2010 followed by a quiescent phase again. During this period, {\sl AF And} has brightened by about 0.44 mag though not much change in temperature was noticed.
\item The duration of primary and secondary outbursts have diverse time scales. While former eruption lasted for more than 4 years, the secondary outburst lasted only for about a year.
\item The spectral analysis of {\sl AF And} exhibits prominent Balmer and He I emission lines along with comparatively weaker FeII and [FeII] emissions. The characteristic P-Cygni profile was also present in this star.
\item Our analysis of this star during the quiescence phase yields a mass-loss rate of about 2.2$\times$10$^{-4}$ M$_{\odot}$ yr$^{-1}$ through stellar winds in the photosphere.
\item We determined the wind terminal velocity for {\sl AF And} between 280$-$300 km~sec$^{-1}$ using P-Cygni profile of HeI emission line.
\end{enumerate}

Although several measurements were made in the past for this star reporting the frequent outburst nevertheless accumulating a multi-band photometric data and monitoring their spectroscopic and photometric nature over a long period of time of our observations and subsequently determination of physical parameters yield some insights into the eruptions towards the late stage of its evolution.

\acknowledgments
We would like to thank referee for the constructive comments that has significantly improved this paper. We are very thankful to the supporting staff at the Nainital and Hoskote for their help in carrying out successful observations for such a long period. KS acknowledges financial support through Raja Ramanna Fellowship. This work has made use of data from the European Space Agency (ESA) mission {\it Gaia} (\url{https://www.cosmos.esa.int/gaia}), processed by the {\it Gaia} Data Processing and Analysis Consortium (DPAC, \url{https://www.cosmos.esa.int/web/gaia/dpac/consortium}). Funding for the DPAC has been provided by national institutions, in particular the institutions participating in the {\it Gaia} Multilateral Agreement. It also makes use of VOSA, developed under the Spanish Virtual Observatory project supported from the Spanish MINECO through grant AyA2017-84089.
\begin{center}
\large{APPENDIX}
\end{center}
M31 has been monitored in our microlensing survey program at Nainital conducted during 1998 to 2002 where LBV star {\sl AF And} serendipitously happens to fall on our target field. Keeping its irregular nature in mind, we made follow-up observations of this star until 2011 of which we present analysis in this paper. \citet{Martin2017a} and \citet{Martin2017b}, however, reported that this star has again gone in the outburst phase between 2015 to 2017 and their follow-up observations are still going on.\\

\bibliographystyle{aasjournal}
\bibliography{ref}
\end{document}

%% file: table01.tex
\begin{table*}
\centering
 \caption{Observing log of the data. First two columns provide the details of the observed night in the form of YMD and mid-Julian day (mJD). Columns 3-5 present the mean psf FWHM, total number of frames observed in $R$ and $I$ band along with corresponding exposure time. In the last 3 columns, we give name of the telescopes, instruments and size of field of view as 2 different telescopes and 3 different CCDs have been used to monitor the target field over a period of around 13 years.} 
  \label{tbl:obs_log}
 \begin{tabular}{cccccccc}
 \tableline  
   Date   &        mJD        &    FWHM   &\multicolumn{2}{c}{No. of frames $\times$ Exp. time (sec)}   &    Tel    &   CCD  & Field of View\\
(yyyymmdd)&      (Days)       & (arcsec)  &     R band  & I band                                      &           &            & (arcmin)  \\ 
\hline
19981121 & 2451139.18552  & 1.87  &  3$\times$1200  &  2$\times$1200  &  ST & 1k$\times$1k & 6.5$\times$6.5  \\
19981122 & 2451140.14350  & 1.50  &  3$\times$1200  &  3$\times$1200  &  ST & 1k$\times$1k & 6.5$\times$6.5  \\
    .    &      .         &  .    &       .         &        .        &         .          & . \\
    .    &      .         &  .    &       .         &        .        &         .          & . \\
19991024 & 2451476.14236  & 1.38  &  2$\times$1200  &        -        &  ST & 2k$\times$2k & 13$\times$13  \\
19991103 & 2451486.09028  & 1.47  &  2$\times$1200  &  3$\times$1200  &  ST & 2k$\times$2k & 13$\times$13  \\
    .    &      .         &  .    &       .         &        .        &         .          & . \\
    .    &      .         &  .    &       .         &        .        &         .          & . \\
20041027 & 2453306.28662  & 2.16  & 47$\times$60    & 30$\times$60    & HCT & 2k$\times$2k & 10$\times$10  \\
20041029 & 2453308.30946  & 1.61  & 23$\times$90    & 10$\times$90    & HCT & 2k$\times$2k & 10$\times$10  \\
    .    &      .         &  .    &       .         &        .        &         .          & . \\
    .    &      .         &  .    &       .         &        .        &         .          & . \\
20101202 & 2455533.23797  & 1.61  &  1$\times$600   &  1$\times$600   &  ST & 2k$\times$2k & 13$\times$13  \\
20111102 & 2455868.14589  & 1.78  &  6$\times$300   &  6$\times$300   &  ST & 2k$\times$2k & 13$\times$13  \\
 \tableline 
 \end{tabular}
 \end{table*}

%% file: table02.tex
 \begin{table}[h]
\centering
 \caption{The $R$ and $I$ magnitudes of the star {\sl AF And} and their corresponding errors in each of the observed nights.}
  \label{tbl:data}
 \begin{tabular}{ccccccc}
 \tableline  
S.N   &       mJD      &    R      &$\sigma_R$&  I       &$\sigma_I$ \\
      &    (2450000+)  &  (mag)    &  (mag)   & (mag)    & (mag)     \\
 \tableline                                                        
  1   &   1139.18552   &  16.018   &  0.003   &    -     &   -       \\
  2   &   1140.14350   &  16.007   &  0.004   &  15.969  &  0.005    \\
  3   &   1143.10731   &  16.011   &  0.003   &  15.994  &  0.008    \\
  4   &   1154.12713   &  16.035   &  0.003   &  16.037  &  0.003    \\
  .   &       .        &    .      &   .      &    .     &   .       \\
  .   &       .        &    .      &   .      &    .     &   .       \\
  .   &       .        &    .      &   .      &    .     &   .       \\
168   &   5533.23797   &  16.781   &  0.010   &  16.809  &  0.020    \\
169   &   5868.14589   &  17.201   &  0.014   &  17.238  &  0.025    \\
 \tableline 
 \end{tabular}
 \end{table}

%% file: table03.tex
\begin{table}
\centering
\caption[caption]{Basic parameters of the {\sl AF And}. $UBVRI$ photometry is taken from the catalogue of \citet{Massey2006}, $JHK_s$ magnitudes from the 2MASS survey \citep{Skrutskie2006}, $W_1W_2W_3W_4$ from the WISE survey \citep{Cutri2012} and $g$-band photometry from the GAIA survey \citep{Gaiamission}.}
\label{tbl:basic_parameters}
\begin{tabular}{ll}
\hline\hline
Parameter               & Value                       \\
RA J2000 (hh:mm:ss)     & $00:43:33.09$               \\
Dec J2000 (dd:mm:ss)    & $+41:12:10.31$              \\
\multicolumn{2}{c}{\underline{Photometric magnitudes from the past studies}}  \\ 
$U$                     & $16.363\pm0.009$            \\
$B$                     & $17.338\pm0.004$            \\
$V$                     & $17.325\pm0.004$            \\
$R$                     & $17.172\pm0.004$            \\
$I$                     & $17.257\pm0.004$            \\
$J$                     & $15.820\pm0.074$            \\
$H$                     & $15.368\pm0.115$            \\
$K_s$                   & $15.406\pm0.188$            \\ 
$g$                     & $17.049\pm0.011$            \\ 
$W1$                    & $14.844\pm0.088$            \\
$W2$                    & $14.992\pm0.150$            \\
$W3$                    & $11.310\pm0.142$            \\
$W4$                    & $7.968\pm0.138$             \\ 
\multicolumn{2}{c}{\underline{Color excess $E(B-V)$}}             \\
\citet{Schlegel1998}    & $0.39\pm0.05$ mag              \\
\citet{Schlafly2011}    & $0.33\pm0.04$ mag             \\
\hline
\end{tabular}
\end{table}

%% file: table04.tex
 \begin{table}[h]
\centering
 \caption{A summary of the magnitude and colour variations during different stages of the evolution of {\sl AF And}.}
  \label{tbl:mag_col_vari}
\small
 \begin{tabular}{cccc}
 \tableline  
    Duration&   $\Delta R$   &   $\Delta I$ &  $\Delta(R-I)$ \\
            &      (mag)     &   (mag)      & (mag)     \\
 \tableline 
Prior to peak brightness     &     0.16     &  0.25  &   0.14   \\
Outburst to Quiescent phase  &     1.36     &  1.50  &   0.26   \\
Quiescent phase              &     0.24     &  0.18  &   0.16   \\
Secondary outburst           &     0.44     &  0.42  &   0.05   \\
 \tableline 
 \end{tabular}
 \end{table}

%% file: table05.tex
\begin{table}[h]
\centering
\caption[caption]{The physical parameters estimated in the present study from the observations of {\sl AF And}.}
\label{tbl:param}
\begin{tabular}{ll}
\hline
Parameter               & Value \\
\hline    
$T_{\textrm{eff}}$ (spectroscopic)    & 33000$\pm$3000\,K \\
$T_{\textrm{eff}}$ (photometric)      & 30700$\pm$2500\,K \\
$M_{bol}$               & -9.49 mag        \\
$t_{1.5}$               &  0.0015 mag~d$^{-1}$        \\
$log(L/L_\odot)$        & 5.70              \\
$v_{terminal}$          & $280-300$ km~s$^{-1}$   \\
Mass loss rate ($\dot{M}$)     & 2.2$\times10^{-4}$ M$_{\odot}$ yr$^{-1}$ \\
Host metallicity [12+log(O/H)] & $8.73-8.89$ dex \\
\hline
\end{tabular}
\end{table}